\documentclass{icrc}

\usepackage{graphicx}
\newcommand{\Dd}[1]{D_{0+}^{#1}}
\newcommand{\fPsi}[3]{\Psi_{#1}^{(#2)}\left(#3\right)}
\newcommand{\sphs}[3]{q_{#1}^{(#2)}\left(#3\right)}
\newcommand{\rr}{\vec{r}}
\begin{document}

\title{Fractional diffusion of cosmic rays}
\author[1]{A.A. Lagutin}
\affil[1]{Altai State University, Barnaul 656099, Russia}
\author[2]{V.V. Uchaikin}
\affil[2]{Ulyanovsk State University, Ulyanovsk 432700, Russia}

\correspondence{A.A. Lagutin (lagutin@theory.dcn-asu.ru)}

\firstpage{1}
\pubyear{2001}


\maketitle

\begin{abstract}

We consider the propagation of galactic cosmic rays under
assumption that the interstellar medium is a fractal one.
An anomalous diffusion equation in terms of fractional
derivatives is used to describe of cosmic ray
propagation. The anomaly in used model results from large
free paths (``L\'evy flights'') of particles between galactic
inhomogeneities and long time a particle stays in a trap.
An asymptotical solution of the
anomalous diffusion equation for point instantaneous and impulse sources
with inverse power spectrum relating to supernova bursts is
found. It covers both subdiffusive and superdiffusive
regimes and is expressed in terms of the stable
distributions. The energy dependence of spectral exponent of
observed particles in different regimes is discussed.
\end{abstract}

\section{Introduction}

A key to understanding the mechanism for cosmic-ray origin and acceleration is
determination how cosmic-ray particles propagate through the interstellar medium
(ISM). If the
propagation process is determined by scattering at magnetic field
inhomogeneities
which have small-scale characters and can be considered as homogeneous Poisson
ensemble, it can be described by a normal diffusion model
(\citet{Ginzburg:1964,Astrophysics:1990}). The normal diffusion process is
characterised by a mean-squared displacement that increases with time,
\(\langle r^2(t)\rangle\propto t\), and by a Gaussian propagator.

However, during a few last decades, many evidences from both theory and
observations of the existence of multiscale structures in the Galaxy have been
found (see, for example, \citet{Lee:1976,Kaplan:1979,Lozinskaya:1986,Ruzmaikin:1988,%
Vanshtein:1989,Bochkarev:1990,Falgaron:1991,Burlaga:1993,Molchanov:1993,%
Meish:1994,Armstrong:1995,Minter:1996,Cadavid:1999}). Filaments, shells,
clouds are entities widely spread in the ISM. A rich variety of structures that
are created in interacting phases having different properties can be related to the
fundamental property of turbulence called intermittency.

The stretching, bending and folding of magnetic field lines by turbulent
motions of a medium partially coupled to the field make magnetic field also highly
intermittent, especially on the smaller scales.
As the turbulent zones do not fill space, the tool of the fractal geometry to
characterise the ISM should be used. In such a fractal-like ISM we certainly do
not expect the normal diffusion to hold.

Generalisation of this equation leads to anomalous diffusion (see
the reviews by \citet{Bouchaud:1990,Isichenko:1992,West:1994} and
\citet{Uchaikin:1999}). In the case of anomalous diffusion
\(\langle r^2(t)\rangle\propto t^\gamma\), \(t\to\infty\), where
the exponent \(\gamma\) differs from 1, a value, that corresponds
to the normal diffusion. Anomalous diffusion has a non-Gaussian
propagator.

In our recent papers (\citet{Lagutin:2000,Lagutin:2001,Lagutin:2001a}) we
proposed an anomalous diffusion (superdiffusion) model for solution of the ``knee'' problem in
primary cosmic-rays spectrum and explanation of different values of
spectral exponent of protons and other nuclei at
$E\sim10^2\div10^5$~GeV/nucleon.  The anomaly results
from large free paths (``L\'evy flights'') of particles between magnetic
domains---traps of the returned type.
These paths are distributed according to inverse power law \(\propto
Ar^{-3-\alpha}\), \(r\to\infty\), \(\alpha\le2\) being an intrinsic
property of fractal structures. We also assumed that the mean time a particle
stays in a trap, \(\langle\tau\rangle\), was finite.

In this paper we consider the propagation of galactic cosmic rays in fractal
ISM without the latter assumption. We suppose now that the particle can spend
anomalously a long time in the trap. An anomalously long time means that
\(\langle\tau\rangle=\int_0^\infty d\tau\,\tau\,q(\tau)=\infty\),
so the distribution of the particles staying in traps, \(q(\tau)\), has a tail of power-law
type \(\propto Bt^{-\beta-1}\), \(t\to\infty\) with \(\beta<1\) (``L\'evy trapping time'').

\section{Particle diffusion in a fractal medium}

After acceleration in a source the particle can be in one of two states: a
state of ``L\'evy flights'' or a state of rest-state of motion in a trap.
Diffusion in this model is a process in which the particle state changes
successively at random moments in time. Based on the continuous time random
walk theory of \citet{Montroll:1965} or on integral equation (see, for
example, \citet{Uchaikin:1999,Uchaikin:1999a}) we can readily derive the fractional diffusion
equation. The equation for Green's function \(G(\vec{r},t,E;E_0)\)
without energy losses and nuclear interactions under condition that
the particle started from origin \(\vec{r_0}=0\) at time \(t_0=0\)
with energy \(E_0\) has the form
\[
\frac{\partial G}{\partial
t}=-D(E,\alpha,\beta)\Dd{1-\beta}(-\Delta)^{\alpha/2} G(\vec{
r},t,E;E_0)
\]
\begin{equation}\label{eq:Green}
+\delta({\rr})\delta(t)\delta(E-E_0).
\end{equation}
Here \(\Dd\mu\) denotes the Riemann-Liouville fractional derivative
(\citet{Samko:1987})
\[
\Dd\mu f(t) \equiv
\frac1{\Gamma(1-\mu)}\frac{d}{dt}\int\limits_0^t (t-\tau)^{-\mu} f(\tau)\,d\tau,
\quad \mu<1,
\]
\(\left( -\Delta \right)^{\alpha/2}\)---the fractional Laplacian (called ``Riss''
operator (\citet{Samko:1987}))
\[
\left( -\Delta \right)^{\alpha/2} f(x) =
\frac{1}{d_{m,l}(\nu)}
\int\limits_{{\rm R}^m} \frac{\Delta_{y}^{l} f(x)}{|y|^{m+\nu}} d y,
\]
where $l > \alpha$, $x \in {\rm R}^m$,  $y \in {\rm R}^m$,
\[
\Delta_{y}^{l} f(x) = \sum_{k=0}^{l} (-1)^{k} {l\choose k} f(x-ky)
\]
and
\[
d_{m,l}(\nu) = \int\limits_{{\rm R}^m} (1 - e^{iy})^l |y|^{-m-\nu} d y.
\]
The anomalous diffusivity \(D(E,\alpha,\beta)\) is determined by the constants
$A$ and $B$ in the asymptotic behaviour for  ``L\'evy flights'' ($A$) and
``L\'evy waiting time'' ($B$) distributions:
\[
D(E,\alpha,\beta)\propto A(\alpha)/B(E,\beta).
\]

Taking into account that the probability to stay in a trap during the time
interval $t$ for particle with charge $Z$ and mass number $A$ depends on
particle rigidity as $\propto R^{-\delta}$, we find
\(D\propto (E/Z)^\delta\).

The solutions of equations (\ref{eq:Green}) with zero boundary conditions at
infinity can be found by Laplace-Fourier transformations
with use of formulae (\citet{Samko:1987})
\[
\int\limits_0^\infty e^{-\lambda t} \Dd\mu G(\vec{ r},t,E;E_0)\,dt
\]
\[= \lambda^\mu \int_0^\infty e^{-\lambda t} G(\vec{ r},t,E;E_0)\,dt =
\lambda^\mu \tilde{G}(\vec{ r},\lambda,E;E_0),
\]
\[
\int\limits_{\rm R^3} e^{i\vec{ k}\vec{ r}} (-\Delta)^{\alpha/2}
G(\vec{ r},t,E;E_0)\, d \vec{ r}
\]
\[=
|\vec{ k}|^\alpha \int\limits_{\rm R^3} e^{i\vec{ k}\vec{ r}}
G(\vec{ r},t,E;E_0)\, d \vec{ r} = |\vec{ k}|^{\alpha}
\tilde{G}(\vec{ k},t,E;E_0).
\]
The  Laplace-Fourier transformation solution of
(\ref{eq:Green}) is
\[
\tilde{G}(\vec{ k},\lambda,E;E_0)=\delta(E-E_0)\lambda^{\beta-1}
\]
\[
\times
\int\limits_0^\infty\exp\left(-[\lambda^\beta+D(E,\alpha,\beta)|\vec{
k}|^\alpha] y\right) dy.
\]
Invert Laplace-Fourier transform we find
\[%
G(\vec{
r},E,t;E_0)=\delta(E-E_0)\bigl(D(E_0,\alpha,\beta)t^\beta\bigr)^{-3/\alpha}
\]
\begin{equation}\label{Green:s}
\times \fPsi3{\alpha,\beta}{|\vec{ r}|\bigl(
D(E_0,\alpha,\beta)t^\beta\bigr)^{-1/\alpha}},
\end{equation}
where
\begin{equation}\label{Psi}
\fPsi3{\alpha,\beta}{r}=\int\limits_0^\infty \sphs3\alpha{r\tau^\beta}
\sphs1{\beta,1}{\tau} \tau^{3\beta/\alpha}\, d\tau.
\end{equation}
Here
\(\sphs3\alpha{r}=(2\pi)^{-3}\int\exp(-ikr-|k|^\alpha)dk\) is the density of three-dimensional
spherically-symmetrical stable distribution with characteristic exponent
\(\alpha\leq2\)~(\citet{Zolotarev:1999,Uchaikin:1999}) and
\(\sphs1{\beta,1}{t}\) is one-sided stable distribution with characteristic
exponent \(\beta\) (\citet{Zolotarev:1986}):
\[
\sphs1{\beta,1}{t}=(2\pi i)^{-1}\int\limits_S\exp(\lambda t-\lambda^\beta)d\lambda.
\]

Let us remember that \(\sphs32r\) is the normal (Gaussian) distribution density,
\(\sphs31r\) is the three-dimensional Cauchy density \([\pi(1+r^2)]^{-2}\),
\(\sphs1{1/2,1}r\) is L\'evy-Smirnov distribution.
Other stable densities cannot be expressed through elementary functions, but
there exist representations in terms of convergent and asymptotic
series~(\citet{Uchaikin:1999}).
Based on equation for Green's function (\ref{eq:Green}) it is easy to
formulate the fractional diffusion equation for concentration:
\[
\frac{\partial N}{\partial t}=-D(E,\alpha,\beta)\Dd{1-\beta}(-\Delta)^{\alpha/2}
 N({\rr},t,E)
\]
\begin{equation}\label{diff}
+S({\rr},t,E),
\end{equation}
where $S(\rr,t,E)$ is a density function of sources distribution.

\section{Spectra}
Using Green's function (\ref{Green:s}) we can find the cosmic ray
concentration for sources interesting for astrophysics. Thus, for point
instantaneous source with inverse power spectrum
\[
S(\vec{r},t,E)=S_0 E^{-p} \delta(\vec{r}) \delta(t)
\]
we have
\[
N(\vec{r},t,E)=S_0 E^{-p} \bigl(D(E,\alpha,\beta)t^\beta\bigr)^{-3/\alpha}
\]
\begin{equation}\label{solve:m}
\times
\fPsi3{\alpha,\beta}{r\bigl(D(E,\alpha,\beta)t^\beta\bigr)^{-1/\alpha}}.
\end{equation}
For point impulse source
\[
S(\rr,t,E)=S'_0 E^{-p} \delta(\rr) \Theta(T-t)\Theta(t)
\]
\[
\Theta(\tau)=\left\{\begin{array}{ll}
1,&\tau>0,\\
0,&\tau<0,\\
\end{array}\right.
\]
the solution is of the form
\[
N(\rr,t,E)=
\frac{S'_0 E^{-p}}{D(E,\alpha,\beta))^{3/\alpha}}\int\limits_{\max[0,t-T]}^t \tau^{-3\beta/\alpha}
\]
\begin{equation}\label{solve:i}
\times
\fPsi3{\alpha,\beta}{|\rr|\bigl(D(E,\alpha,\beta)\tau^\beta\bigr)^{-1/\alpha}}.
\end{equation}

Using the representation \(N=N_0E^{-\eta}\) and the asymptotic behaviour of
the scaling function \(\fPsi{}{\alpha,\beta}{r}\) (\(\alpha<2\), \(\beta<1\))
\[
\fPsi3{\alpha,\beta}{r\to0} \propto r^{-(3-\alpha)},
\]
\[
\fPsi3{\alpha,\beta}{r\to\infty} \propto r^{-3-\alpha},
\]
one can evaluate the variation of spectral exponent
\(\Delta\eta=\eta(E\to\infty)-\eta(E\to0)\). It follows from (\ref{solve:m}) that
\[
N(\rr,t,E) \propto E^{-p+\delta},\quad E\to0,
\]
\[
N(\rr,t,E) \propto E^{-p-\delta},\quad E\to\infty,
\]
\[
\Delta \eta=2\delta.
\]
In other words, in case of point instantaneous source in both subdiffusive
(\(\beta<\alpha/2)\) and superdiffusive (\(\beta>\alpha/2)\) regimes the
spectral exponent of observed particle increases with energy on \(2\delta\),
i.e. the cosmic ray spectrum steepens (the ``knee'').

The similar estimates for the impulse source (see (\ref{solve:i})) give
\[
N(\rr,t,E) \propto E^{-p+\delta},\quad E\to0,
\]
\[
N(\rr,t,E) \propto E^{-p-\delta/\beta},\quad E\to\infty,
\]
\[
\Delta \eta=\delta(1+1/\beta).
\]
Our analytical and numerical studies however show that this property of
energy spectrum (the ``knee'') is lacking in the regimes of normal diffusion
(\(\alpha=2\), \(\beta=1\)) and subdiffusion (\(\alpha=2\), \(\beta<1\)).

\balance
\section{Conclusion}
We considered the propagation of galactic cosmic ray in the fractal interstellar
medium. Anomalous diffusion equation in terms of fractional derivates
describing of cosmic ray propagation has been formulated.
An asymptotical solution of
this equation covered both subdiffusive and superdiffusive regimes has been
expressed in terms of stable distributions.

Our results showed that the ``knee'' in the primary cosmic ray spectrum is due
to anomalously large free paths (``L\'evy flights'') of particles, being an
intrinsic property of the fractal interstellar medium.

This work was supported by RFBR grants 0001 00284 and 0002 17507 and program
``Integration'' (project 2.1--252).



\begin{thebibliography}{99}
\bibitem[Armstrong et al.(1995)]{Armstrong:1995}
Armstrong J.W., Rickett B.J., Spangler S.R. Ap.J. 443, 209--221, 1995.
\bibitem[Berezinsky et al. (1990)]{Astrophysics:1990}
Berezinsky V.S., Bulanov S.V., Ginzburg V.L. et al. Astrophysics of cosmic
rays. North Holland, Amsterdam, 1990.
\bibitem[Bochkarev (1990)]{Bochkarev:1990}
Bochkarev N.G. The local interstellar medium. M.: Nauka, 1990.
\bibitem[Bouchaud and Georges (1990)]{Bouchaud:1990}
Bouchaud J.-P., Georges A. Phys.Rep. 195, 127--293, 1990.
\bibitem[Burlaga et al. (1993)]{Burlaga:1993}
Burlaga L.F., Perko J., Pirraglia J. Ap.J. 407, 347--358, 1993.
\bibitem[Cadavid et al. (1999)]{Cadavid:1999}
Cadavid A.C., Lawrence J.K., Ruzmaikin A.A.  Ap.J. 521, 844--850, 1999.
\bibitem[Falgarone et al. (1991)]{Falgaron:1991}
Falgarone E., Phillips T.G., Walker C.K. Ap.J. 378, 186--201, 1991.
\bibitem[Ginzburg and Syrovatskii (1964)]{Ginzburg:1964}
Ginzburg V.L. and Syrovatskii S.I. Origin of cosmic rays. Pergamon Press, 1964.
\bibitem[Isichenko (1992)]{Isichenko:1992}
Isichenko M.B. Rev.Mod.Phys. 64, 961--1043, 1992.
\bibitem[Kaplan and Pikelner (1979)]{Kaplan:1979}
Kaplan S.A., Pikelner S.B. Physics of the interstellar medium. M.:Nauka, 1979.
\bibitem[Lagutin et al. (2000)]{Lagutin:2000}
Lagutin A.A., Nikulin Yu.A. Uchaikin V.V. Preprint ASU--2000/4, Barnaul,
2000.
\bibitem[Lagutin et al. (2001a)]{Lagutin:2001}
Lagutin A.A., Nikulin Yu.A. Uchaikin V.V. Nucl.Phys.B, 97, 267--270, 2001.
\bibitem[Lagutin et al. (2001b)]{Lagutin:2001a}
Lagutin A.A., Nikulin Yu.A. Uchaikin V.V. Izv.RAN. Ser.fiz., 2001 (to be published).
\bibitem[Lee and Jokipii (1976)]{Lee:1976}
Lee L.C., Jokipii J.R. Ap.J. 206, 735--743, 1976.
\bibitem[Lozinskaya (1986)]{Lozinskaya:1986}
Lozinskaya T.A. Supernova and star wind: interactions with galactic gas. M.:Nauka, 1986.
\bibitem[Meish and Bally (1994)]{Meish:1994}
Meish M.S. and Bally J. Ap.J. 429, 645--671, 1994.
\bibitem[Minter and Spangler (1996)]{Minter:1996}
Minter A.H. and Spangler S.R. Ap.J. 458, 194--214, 1996.
\bibitem[Molchanov et al.(1993)]{Molchanov:1993}
Molchanov S.A., Ruzmaikin A.A., Sokolov D.D.  In book Nonlinear waves:
physics and astrophysics. M.:Nauka, 1993, 47.
\bibitem[Montroll and Weiss (1965)]{Montroll:1965}
Montroll E.W., Weiss G.H. J. Math. Phys.  6, 167--181, 1965.
\bibitem[Ruzmaikin et al. (1988)]{Ruzmaikin:1988}
Ruzmaikin A.A., Sokolov D.D., Shukurov A.M. Magnetic fields of Galaxies.
Kluwer, Dordrecht, 1988.
\bibitem[Samko et al. (1987)]{Samko:1987}
Samko S.G., Kilbas A.A., Marichev O.I. Fractional integrals and derivations
and some applications (in Russian). Minsk: Nauka, 1987.
\bibitem[Uchaikin (1999)]{Uchaikin:1999a}
Uchaikin V.V. JETP 88,  1155--1163, 1999.
\bibitem[Uchaikin and Zolotarev (1999)]{Uchaikin:1999}
Uchaikin V.V., Zolotarev V.M. Chance and Stability. VSP, Netherlands,
Utrecht, 1999.
\bibitem[Vanshtein et al. (1989)]{Vanshtein:1989}
Vainshtein S.I., Bykov A.M., Toptygin I.N. Turbulence,
stream layers and shock wave in cosmic plasm.
M.:Nauka, 1989 (in Russian).
\bibitem[West and Deering (1994)]{West:1994}
West B.J. and Deering W. Phys.Rep. 246, 1--100, 1994.
\bibitem[Zolotarev et al. (1999)]{Zolotarev:1999}
Zolotarev V.M., Uchaikin V.V., Saenko V.V. ZhETF 115, 1411--1425, 1999.
\bibitem[Zolotarev (1983)]{Zolotarev:1986}
Zolotarev V.M. One-dimensional stable laws. M.: Nauka, 1983.



\end{thebibliography}
\end{document}